\documentclass[a4paper,11pt]{article}
\pdfoutput=1 

\usepackage{jcappub} 

\usepackage[T1]{fontenc} 

\usepackage{graphicx}
\usepackage{dcolumn}
\usepackage{bm}
\usepackage{epsfig}
\usepackage{graphicx}
\usepackage{hyperref}
\usepackage[usenames]{color}
\usepackage{url}
\usepackage{enumitem}
\usepackage{float}
\usepackage{longtable,lscape}

\hypersetup{
    colorlinks=true,
    linkcolor=red,
    citecolor=blue,
}

\newcommand{\rv}{\textcolor{black}}

\newcommand{\remove}[1]{}

\newcommand{\hoallthree}{$  H_0=73.04  \pm  1.04  $ km s$^{-1}$ Mpc$^{-1}$}

\def\etal{{\frenchspacing\it et al.}}
\def\ie{{\frenchspacing\it i.e.}}

\def\be{\begin{equation}}
\def\ee{\end{equation}}
\def\ba{\begin{eqnarray}}
\def\ea{\end{eqnarray}}
\title{Data-driven and Almost Model-independent Reconstruction of Modified Gravity}

\author[a]{Yuhao Mu,}
\author[b]{En-Kun Li,}

\author[a,1]{Lixin Xu \note{Corresponding author.}}

\affiliation[a]{Institute of Theoretical Physics, School of Physics, Dalian University of Technology, Dalian, 116024, P. R. China}
\affiliation[b]{MOE Key Laboratory of TianQin Mission, TianQin Research Center for Gravitational Physics \& School of Physics and Astronomy, Frontiers Science Center for TianQin, Gravitational Wave Research Center of CNSA, Sun Yat-sen University (Zhuhai Campus), Zhuhai 519082, China}

\emailAdd{lxxu@dlut.edu.cn}

\abstract{
 
In this paper, a modified factor $\mu$, which characterizes modified gravity in the linear matter density perturbation theory, is reconstructed in a data-driven and almost model-independent way via Gaussian process by using currently available cosmic observations. Utilizing the Pantheon+ SNe Ia samples, the observed Hubble parameter $H(z)$ and the redshift space distortion $f\sigma_8(z)$ data points, one finds out a time varying $\mu$ at low redshifts. The reconstructed $\mu$ implies that more complicated modified gravity beyond the simplest general relativity and the Dvali-Gabadadze-Porrati braneworld model is required.  
    
}

\begin{document}

\maketitle
\flushbottom

\section{Introduction}\label{sec:intro}

It was well known that our Universe is undergoing an accelerated expansion \cite{ref:Riess98,ref:Perlmuter99}. Although a convincing explanation to this accelerated expansion phase is still empty, one may reach to a consensus that new physics beyond `standard model' might be demanded. As of today, two approaches to address this accelerated expanding Universe have been developed in the last twenty years, please see Refs. \cite{ref:DEReview1,ref:DEReview2,ref:DEReview3,ref:DEReview4,ref:DEReview5,ref:DEReview6,ref:DEReview7,ref:MG1,ref:MG2,ref:MG3,Ishak:2018his,Amendola:2016saw,Joyce:2014kja,Nojiri:2010wj,Nojiri:2017ncd}  for comprehensive reviews but not for a complete list. One is to introduce a new energy component having a negative pressure, named as dark energy (DE). Another is 
to modify the general relativity (GR) at large scales, the so-called modified gravity theory (MG). In phenomenon, both DE and MG can provide a same evolution history of our Universe \cite{ref:background}. But the growth of the large scale structure under different scenarios might be totally different even they share the same expansion history. Thus one advocates to use the growth rate $f\equiv d\ln\delta/d\ln a$ to discriminate DE from MG \cite{Peebles,Fry,Lightman,wang,lue,Aquaviva,gong08b,polarski,linder,Koyama,Koivisto,Daniel,knox,ishak2006,laszlo,Zhang,Hu,ref:Ishak,ref:Xu2013}, where  $\delta(z)\equiv\delta\rho_m/\rho_m$ is the linear matter density contrast, $a=1/(1+z)$ is the scale factor. In fact, for MG theory the growth rate is genernally given in terms of the redshift $z$ as
 \be
(1+z)f'-f^2+\left[(1+z)\frac{H'}{H}-2\right]f+\frac{3\mu}{2}\Omega_m=0,\label{eq:fgrowthz}
\ee
where $\Omega_{m}=H^2_0\Omega_{m0}(1+z)^{3}/H^2$ is the dimensionless dark matter energy density, $H$ is the Hubble parameter and the prime $'$ denotes the derivative with respect to the redshift $z$. The function $\mu$
\be
\mu=\frac{G_{\rm eff}}{G_{\rm N}},
\ee
effectively corresponds to a modification to the Newtonian constant $G_{\rm N}$ and characterizes MG theory at the linear matter density perturbation level. $\mu\equiv 1$ is respected in GR. In general, this modified factor $\mu$ might be a function of time and space, i.e a function of the redshift $z$ and the scale $k$ in Fourier space $\mu(z,k)$ \cite{ref:Xuprd2015}. For example in $f(R)$ gravity, $\mu$ is given as \cite{ref:MG3}
\be
\mu=\frac{1}{F}\frac{4+3M^2a^2/k^2}{3(1+M^2a^2/k^2)},
\ee   
where $R$ is the Ricci scalar, $F=d f(R)/dR=f_{,R}$, $M^2=R(1/m-1)/3$ and $m=Rf_{,RR}/f_{,R}$. However for the Dvali-Gabadadze-Porrati (DGP) braneworld model \cite{ref:DGP}, the scale $k$ free $\mu$ is given as \cite{lue,Koyama} 
\be
\mu=1+\frac{1}{3\beta},~~~\beta\equiv 1-2Hr_c\left(1+\frac{\dot{H}}{3H^2}\right),
\ee
where $r_c\equiv \kappa^2_{(5)}/2\kappa^2_{(4)}$ is the length scale related to the gravitational coupling constant in five- and four-dimensional space-time. For $f(T)$ gravity, $\mu$ is also scale $k$ free and is written as \cite{ref:ZhengHuangfT2011}
\be
\mu=\frac{1}{1+f_{T}},\label{eq:mufT}
\ee
where $f_{T}=df(T)/dT$. While a general parametrized form of $\mu$ can be given as \cite{ref:Amendolajcap2008,ref:BZ2008,ref:Pogosianprd2010,ref:MGCAMB,ref:XuMGGW2015} 
\be
\mu(k,a)=\frac{1+\beta_1\lambda^2_1k^2a^s}{1+\lambda^2_1k^2a^s},
\ee 
where $\beta_1$ is dimensionless coupling and $\lambda_1$ have dimension of length, and in the other form as \cite{ref:FeliceKobayashiTsujikawa2011,ref:Solomon2014,ref:Konnig2014}
\be
\mu(k,a)=\frac{1+p_1(a)k^2}{p_2(a)+p_3(a)k^2},
\ee
where $p_i(a)$'s are functions of the scale factor $a$. Thus, the detection of any deviation from $\mu\equiv 1$ implies a possible modification to GR. With the above observations, one can take the modified factor $\mu$ from Eq. \eqref{eq:fgrowthz} 
\be
\mu =\frac{2}{3\Omega_m} \left\{-(1+z)f'+f^2 +\left[2 - (1+z) \frac{H'}{H} \right] f \right\},    \label{eq:mu}
\ee       
as a useful indicator to MG. It would be more interesting, the modified factor $\mu$ can be reconstructed in a data-driven and model-independent way, \ie without assuming any peculiar parameterized form of $\mu$. This is the main motivation and task of this work.  

It is fortunate that the observed values of the growth rate $f_{\rm obs}=\beta b$ can be derived from the redshift space distortion (RSD) parameter $\beta(z)$ and the linear bias $b(z)$, where a particular fiducial $\Lambda$CDM model is used \cite{ref:weakness}. But it cannot be used directly, because the measurements of the linear growth rate are degenerate to the bias $b$ or clustering amplitude in the power spectra. To remedy this weak point, Song and Percival proposed to use $f\sigma_8(z)$, which is almost model-independent and provides a good test for dark energy models even without the knowledge of the bias or $\sigma_8$ \cite{ref:Song}, where $\sigma_8(z)=\sigma_8(z=0)\delta(z)/\delta(z=0)=\sigma_{8,0}\delta(z)/\delta_0$ within spheres of radius $8h^{-1}$Mpc, and the subscript ``0'' indicates the present value of the corresponding quantity. And we already have $63$ $f\sigma_{8,\rm obs}$ data points, please see Table A1 in Ref. \cite{ref:Li2021} and references therein for example. Meanwhile to remove the dependence to a particular fiducial $\Lambda$CDM model, the Alcock-Paczynski (AP) effect \cite{Alcock:1979mp} should also be considered. 

In order to reconstruct the Hubble parameter $H(z)$ and $H'(z)$ (the first derivative of $H(z)$ with respect to the redshift $z$), we use the the distance moduli from Pantheon+ SNe Ia samples \cite{Scolnic:2021,Brout:2022} and the observational Hubble data points from the cosmic chronometers (CC) and from clustering measurements (BAO), see Table A2 and A3 in Ref. \cite{ref:Li2021} and references therein for example.

For the reconstructed functions, we resort to Gaussian process \cite{Rasmussen:2006,GaPP:2012}, which was used extensively in cosmology in the last few years \cite{GaPP:2012,Holsclaw:2010prl,Holsclaw:2010prd,Santos-da-Costa:2015,Shafieloo:2012,Yahya:2014,Yang:2015,Cai:2016,Zhang:2016,Cai:2015,Wang:2017,ref:Wei2019,ref:Liang2022,ref:Mu2023,ref:gapptelep2021,ref:gappH0Colgain2021,ref:gappCalderon2022,ref:gappLi2022,ref:gappCalderon2023}. Gaussian process can reconstruct the function $g(x)$ from data points $g(x_i)\pm\sigma_i$ via a point-to-point Gaussian distribution \cite{Rasmussen:2006,GaPP:2012}, without assuming a specific parameterized form.

This paper is structured as follows. In the next Section \ref{sec:cg}, we present the main methodology. In Section \ref{sec:observation}, the modified factor $\mu$ is reconstructed in a data-driven and almost model-independent way, but it is only sticked to the case where $\mu$ varies with respect to the redshift $z$, mainly due to the scale-dependence is outside the range probed by large scale structure surveys within the linear matter perturbation theory. The Section \ref{sec:con} is the conclusion and discussion.

\section{Methodology} \label{sec:cg}

\rv{Without assuming a specific parameterized form, Gaussian process can reconstruct the function $g(x)$ from data points $g(x_i)\pm\sigma_i$ via a point-to-point Gaussian distribution \cite{Rasmussen:2006,GaPP:2012}, where the expected value $\bar{g}$ and the variance $\sigma_g^2$ of the function $g(x)$ for $N$ number of data points are given by
\ba
\bar{g}(x)&=&\sum_{i,j=1}^Nk(x,x_i)(M^{-1})_{ij}g(x_j),\label{eq:mux}\\
\sigma_g^2(x)&=&k(x,x)-\sum_{i,j=1}^Nk(x,x_i)(M^{-1})_{ij}k(x_j,x),\label{eq:varx}
\ea
where $M_{ij}=k(x_i,x_j)+C_{ij}$ is the covariance matrix, and $C_{ij}$ is the covariance matrix of the data points, and $k(x, \tilde{x})$ is the covariance function or kernel between the points $x$ and $\tilde{x}$, which is usually taken as the squared exponential covariance function in the form
\be
k(x,\tilde{x}) =
\sigma_g^2 \exp\left[-\frac{(x - \tilde{x})^2}{2\ell^2} \right],\label{eq:kxx}
\ee
where the `hyper-parameter' $\sigma_g$ denotes the typical change in the $y$-direction, and $\ell$ characterizes the distance traveling in $x$-direction to get a significant change in a function. These two `hyper-parameters' $\sigma_g$ and $\ell$ are determined in Gaussian process by maximizing the logarithmic marginalized likelihood function  
 \be
\ln\mathcal{L}=-\frac{1}{2}\sum_{i,j=1}^N g(x_i)\left(M^{-1}\right)_{ij}g(x_j)-\frac{1}{2}\ln|M|-\frac{1}{2}N\ln2\pi,\label{eq:likelihood}
\ee
where $|M|$ is the determinant of $M_{ij}$. As shown in Ref. \cite{GaPP:2012}, for a set of input points $\bm X=\{x_i\}$, the covariance matrix $[K(\bm X,\bm X)]_{ij}=k(x_i,x_j)$  can generate a random (quite arbitrary) function $g(\bm X)$ at $\bm X^\ast$ from the Gaussian process via $g^{\ast}\sim \mathcal{N}(\bar{g}^{\ast},K(\bm X^\ast,\bm X^\ast))$, since the function is not restricted by any observations, where $\bar{g}^{\ast}$ is the {\it a priori} assumed mean of $g^\ast=g(\bm X^\ast)$, where $\mathcal{N}$ means the Gaussian process $\mathcal{GP}$ is evaluated at specific points $\bm X^\ast$, and $g(\bm X^\ast)$ is a random value draw from a normal distribution. In this situation the hyperparameters have not been trained yet, thus the scale of the $y$-axis is not fixed and all functions are still possible. Adding the observed data constrains the function space as illustrated in the Figure 2. of Ref. \cite{GaPP:2012}. Reconstructions of higher derivatives can be done analogously, for example the first derivative (here the prime denotes the derivative with respect to the variable $x$) is given as
\ba
\bar{g'}(x)&=&\sum_{i,j=1}^N k'(x,x_i)(M^{-1})_{ij}g(x_j),\label{eq:muxprime}\\
\sigma_{g'}^{2}(x)&=&k''(x,x)-\sum_{i,j=1}^N k'(x,x_i)(M^{-1})_{ij}k'(x_j,x),\label{eq:varxprime}
\ea
where
\ba
k'(x,x_i)&=&\frac{\partial k(x,x_i)}{\partial x_i},\\
k''(x_i,x_j)&=&\frac{\partial^2 k(x_i,x_j)}{\partial x_i\partial x_j}.
\ea
The hyperparameters are trained in the same way as that for the reconstruction of $g(x)$, because the marginal likelihood Eq. (\ref{eq:likelihood}) depends only on the observations not on the reconstructed function one wants to reconstruct  \cite{GaPP:2012}. In this work, we will modify and use the publicly available {\bf GaPP} code \footnote{\url{https://github.com/carlosandrepaes/GaPP}.} \cite{GaPP:2012}.}

Starting from the definition of the growth rate
 \be
 f\equiv\frac{d\ln\delta}{d\ln a}
 =-(1+z)\frac{\delta^\prime}{\delta}=-(1+z)\frac{\delta^\prime/\delta_0}{\delta/\delta_0},\label{eq:fdelta}
 \ee
 one has the first order derivative of $f(z)$ with respect to the redshift $z$
 \be
 f'=-\frac{\delta^\prime(z)/\delta_0}{\delta(z)/\delta_0}-(1+z)\left[ \frac{\delta''(z)/\delta_0}{\delta(z)/\delta_0} - \frac{(\delta'(z)/\delta_0)^2}{(\delta(z)/\delta_0)^2}\right],\label{eq:fprime}
 \ee
where
 \ba
 \frac{\delta^\prime(z)}{\delta_0}&=&
 -\frac{1}{\sigma_{8,\,0}}\frac{f\sigma_8(z)}{1+z}\,,\label{eq3}\\
 \frac{\delta(z)}{\delta_0} &=& 1-\frac{1}{\sigma_{8,\,0}} \int_0^z
 \frac{f\sigma_8(\tilde{z})}{1+\tilde{z}}\,d\tilde{z},\label{eq:deltaodelta0}\\
 \frac{\delta''(z)}{\delta_0}&\equiv& \left[ \frac{\delta^\prime(z)}{\delta_0}\right]'= \frac{1}{\sigma_{8,\,0}}\frac{f\sigma_8(z)}{(1+z)^2}- \frac{1}{\sigma_{8,\,0}}\frac{[f\sigma_8(z)]'}{1+z}.
 \ea
Therefore, one can reconstruct the relevant functions $f$ and $f'$ by using the observed $f\sigma_{8,\rm obs}$ data via Gaussian processes. For the $63$ $f\sigma_{8,\,obs}$ data points, please see Table A1 in Ref. \cite{ref:Li2021} and references therein for example. The {\it Planck}
 2018 results $\sigma_{8,0}=0.8111\pm 0.0060$ \cite{Aghanim:2018eyx,Akrami:2018vks} is adopted in the last step to obtain $f(z)$ and $f'(z)$. \rv{The uncertainty of $\sigma_{8,0}$ is included to the final reconstructed functions $f$ and $f'$ by the error propagation.}

We take $D_C(z)$, the comoving distance for a spatially flat Universe
\be 
D_C(z)=c\int_0^{z}\frac{d\tilde{z}}{H(\tilde{z})},\label{eq:chi}
\ee  
as a new observable, where $c$ is the speed of light. The corresponding covariance matrix for $D_C(z)$ can be given as 
\be
C^{\rm tot}_{ij} =\left[\frac{D^i_L}{(1+z_i)^2}\right]^2\sigma^2_{z_{i}}\delta_{ij}+\frac{\ln10 D^i_L}{5(1+z_i)} \tilde{C}^{\rm tot}_{ij}\frac{\ln10 D^j_L}{5(1+z_j)} , \label{eq:covariance}
\ee 
where  $D_L=(1+z)D_C$ is luminosity distance. This covariance matrix can be derived by error propagation equation, here $z_i$ and $D^i_L$ are the redshift and the observed luminosity distance of the $i$-th SNe Ia respectively, and $\sigma_{z_{i}}$ is the $1\sigma$ error for $z_i$. And $\delta_{ij}$ is the standard Kronecker symbol. $\tilde{C}^{\rm tot}_{ij}$ in the last term is total distance modulus covariance matrix for Pantheon+ SN Ia samples \footnote{The data points are available online \url{https://github.com/PantheonPlusSH0ES/DataRelease}.} \cite{Scolnic:2021,Brout:2022}, and there is no Einstein's summation convention. This variance $C^{\rm tot}_{ij}$ will be added to the covariance matrix
$K(\bm X,\bm X) + C^{\rm tot}$ where $[K(\bm X,\bm X)]_{ij}=k(x_i,x_j)$ is the covariance matrix for a set of input points $\bm X=\{x_i\}$. Similarly, in order to reconstruct $D'_C$ from CC+BAO, the following covariance matrix is needed
\be
C^{H}_{ij} = \left[\frac{c}{H_i^2}\right]^2\sigma^2_{H_i} \delta_{ij}. \label{eq:Hcovariance}
\ee  
In this manner, $D_C(z)$ is reconstructed by jointed combination from SNe Ia distance moduli and the observation Hubble data points. Then the Hubble parameter $H(z)$ via the relations $H(z)=c/D'_C(z)$ can be obtained. Thus the currently best reconstructed $H(z)$ is presented in a data-driven and cosmological model-independent way.  
   
In order to reconstruct $f(z)$, we should treat the covariance matrix and $f\sigma_{8,\rm obs}$ carefully. At first, the covariance matrix for the $63$ $f\sigma_{8,\rm obs}$ data points are assumed to be diagonal, but with the exception to the WiggleZ subset of the data (three data points):
\be
C_{i,j}^{\text{WiggleZ}} = 10^{-3} \times
\begin{pmatrix}
6.400 & 2.570 & 0.000 \\
2.570 & 3.969 & 2.540 \\
0.000 & 2.540 & 5.184
\end{pmatrix}. \label{eq:Cij_WiggleZ}
\ee
Secondly, all the $f\sigma_{8,\rm obs}(z)$ data are obtained assuming a fiducial $\Lambda$CDM cosmology.
Thus to make data analysis consistent, the Alcock-Paczynski (AP) effect \cite{Alcock:1979mp} should be considered.
In this work, we will use the following rough approximation of the AP effect \cite{Macaulay:2013swa,Kazantzidis:2018rnb}
\be
f\sigma_{8,\text{ap}}(z) \simeq \frac{H(z) D_A(z)}{H^{\text{fid}}(z,\Omega_m) D_A^{\text{fid}}(z,\Omega_m) } f\sigma_{8,\text{obs}}(z),
\label{eq:AP}
\ee
where $D_A(z)$ is the angular diameter distance
\be
D_A(z) = \frac{c}{1+z} \int_0^z \frac{d\tilde{z}}{H(\tilde{z})},
\label{eq:D_A}
\ee
for the spatially flat Universe. Thirdly, to make the whole analysis consistent, data-driven and cosmological model-independent, the functions $H(z)$ and $D_A(z)$ in Eq. (\ref{eq:AP}) for fiducial model correction respect to the reconstructed results from the Pantheon+ SNe Ia samples and CC+BAO data points, where the relations $H(z)=c/D'_C(z)$ and $D_A(z)=D_C(z)/(1+z)$ are used. In this way, the central value of $f\sigma_{8,\text{ap}}(z)$ and its covariance matrix can be calculated according to Eq. (\ref{eq:AP}). Considering AP effect, the covariance matrix for $f\sigma_{8,\text{obs}}(z)$ is given as
\begin{equation}
\text{Cov}_{ij}^{\rm ap} = q_i q_j \text{Cov}_{ij}^* + p_i p_j \text{Cov}_{ij}^{\rm HD},
\label{eq:cov_ap}
\end{equation}
where $\text{Cov}^*$ and $\text{Cov}^{\rm HD}$ are the covariance of observational RSD data and the reconstructed $H(z)D_A(z)$, respectively, $q_i = H_i D_{A,i}/(H_i^{\text{fid}} D_{A,i}^{\rm fid})$, and $p_i = f\sigma_{8,i,\text{obs}}/(H_i^{\text{fid}} D_{A,i}^{\rm fid})$.

\section{Reconstructed Modified Factor $\mu$ via Gaussian Process} \label{sec:observation}

In this section, we will present the reconstructed modified factor $\mu$ from cosmic observations via Gaussian process. To obtain $H(z)$ and $H'(z)$, we firstly reconstruct the comoving distance $D_C(z)$ and $D'_C(z)$ by using the distance moduli from Pantheon+ SNe Ia samples and observational Hubble parameters from CC+BAO. As results, one has the expansion history $H(z)$ and $\Omega_{m}(z)$ consequently without assuming any cosmological model. Then the growth rate $f(z)$ will be obtained by reconstruction from $f\sigma_{8,\rm obs}$ data after implementing the Alcock-Paczynski (AP) effect \cite{Alcock:1979mp} correction to guarantee the consistence.  

For the SNe Ia data points, we use the recently released Pantheon+ samples which consist of 1701 light curves of 1550 spectroscopically confirmed SNe Ia coming from 18 different sky surveys. In this sample, the redshifts range from $z=0.00122$ to $2.26137$. As pointed as in our previous study \cite{ref:cosmography2011}, due to the degeneracy between $H_0$ and the absolute magnitude $M$, the SNe Ia cannot give any prediction of $H_0$ value without calibration. In using the measurement of \hoallthree from SH0ES, and making it consistent and free of redundancy, some Pantheon+ SNe Ia data points (marked as {\bf USED\_IN\_SH0ES\_HF=1}) are removed where they were already used in the Hubble flow dataset \cite{Riess:2022}.  

For the observational Hubble data points, we use the $31$ so-called cosmic chronometers (CC) $H(z)$ data points which are determined by computing the age difference $\Delta t$ between passively-evolving galaxies at close redshifts, and the $23$ $H(z)$ data points from clustering measurements (BAO), see Table A2 and A3 in Ref. \cite{ref:Li2021} and references therein for the lists. Although some of the $H(z)$ data points from clustering measurements are correlated due to the overlap between the galaxy samples, we assume that they are independent measurements for simplicity.   

\begin{figure*}[tbp]  
\includegraphics[scale=0.5]{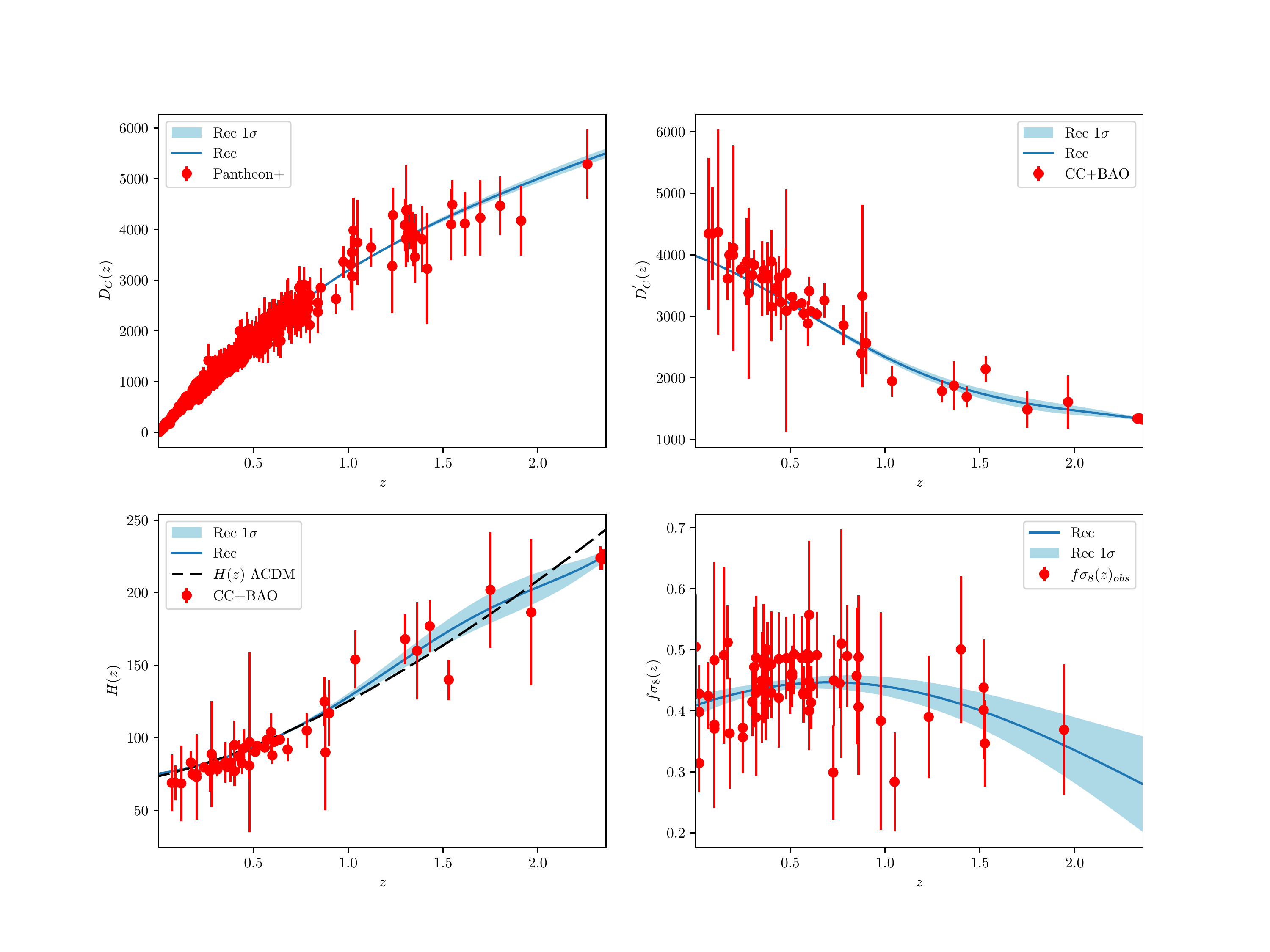}
\caption{Upper left panel: The reconstructed comoving distance $D_C(z)$ with $1\sigma$ regions and Pantheon+ samples denoted as error bar lines; Upper right panel: The reconstructed $D'_C(z)$ with $1\sigma$ regions are shown, where the observational data point from CC+ABO are plotted as error bar lines; Lower left panel: The reconstructed Hubble diagram $H(z)$ with $1\sigma$ shadow regions are plotted, where the observed Hubble parameter values are plotted as error bar lines; Lower right panel: The reconstructed $f\sigma_8(z)$ with $1\sigma$ regions are plotted, where the observed $f\sigma_{8,\rm obs}$ are plotted as error bar lines.}\label{fig:DCS}
\end{figure*}

Implementing Gaussian process, the comoving distance $D_C(z)$ and $D'_C(z)$ are reconstructed as shown in the upper left and right panels respectively in Figure \ref{fig:DCS} where the $1\sigma$ uncertainty regions are also plotted as shadows. It can be seen that narrow uncertainty is obtained in the reconstructed $D_C(z)$, it is mainly due to the addition of CC+BAO data points as an extra constraint to $D'_C(z)$. While a large error regions appear at high redshifts due to the sparse data points. The Hubble parameter $H(z)$ and $f\sigma_8(z)$ with $1\sigma$ shadow regions are also plotted in the lower left and right panels of Figure \ref{fig:DCS} respectively. In the Hubble diagram, a spatially flat $\Lambda$CDM cosmology, i.e. $H^2(z)=H^2_0[\Omega_{m0}(1+z)^3+\Omega_{\Lambda0}]$ with $\Omega_{m0}=0.334$  ($\Omega_{\Lambda0}=1-\Omega_{m0}$) from SH0ES \cite{Riess:2022} is plotted as dark dashed line for comparison. It shows that the suppression of $H(z)$ values at high redshifts comes from the lower observed values as to that predicted from the $\Lambda$CDM cosmology. The observed data points are also included as error bar lines in the corresponding panels. It can be seen that the reconstructed functions match the observed values really well.     

\begin{figure*}[tbp]  
\includegraphics[scale=0.5]{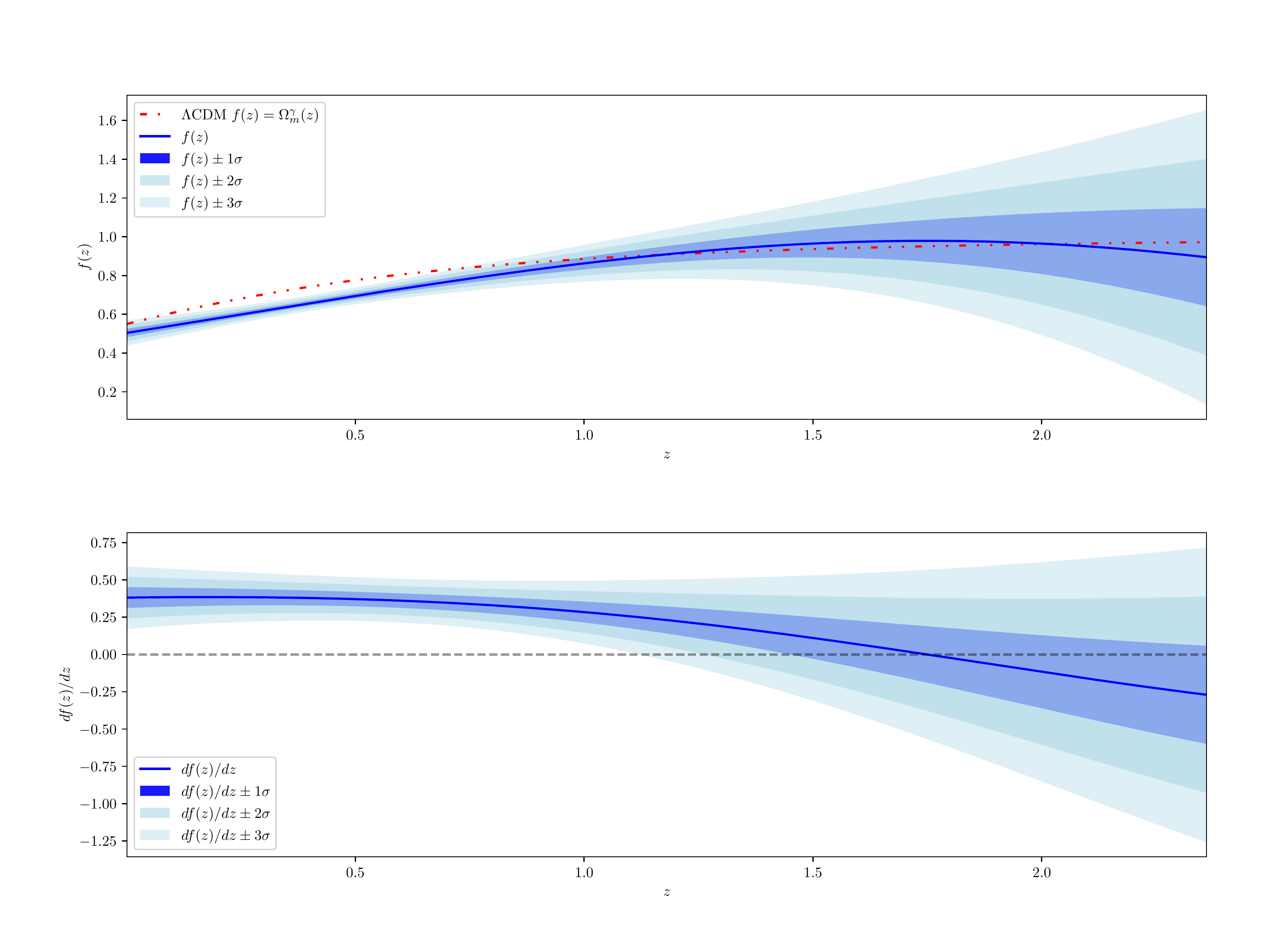}
\caption{Upper Panel: The reconstructed growth rate $f(z)$ and its $1-3\sigma$ regions, where $f(z)=\Omega^{\gamma}_{m}(z)$ for the spatially flat $\Lambda$CDM cosmology is plotted as red dashed line for comparison. Lower Panel: The corresponding reconstructed $f'(z)$, the first derivative of $f(z)$ with respect to the redshift $z$, which can be obtained from $f(z)$ numerically or derived from the reconstructed $f\sigma_8(z)$ via Eq. (\ref{eq:fprime}). We have checked that two approaches give the same results.} \label{fig:fgrowth}
\end{figure*}

Correspondingly, the reconstructed growth rate $f(z)$ and $f'(z)$ (the first derivative of $f(z)$ with respect to the redshift $z$), are plotted in Figure \ref{fig:fgrowth}. In the upper panel of Figure \ref{fig:fgrowth}, the reconstructed growth rate $f(z)$ with $1-3\sigma$ regions are shown, where $f(z)=\Omega^{\gamma}_{m}(z)$ for the spatially flat $\Lambda$CDM cosmology is plotted as red dashed line for comparison, here the standard $\gamma=6/11$ is adopted. In the lower panel of Figure \ref{fig:fgrowth}, the corresponding reconstructed $f'(z)$ with $1-3\sigma$ regions are plotted, which can be obtained from $f(z)$ by calculating $f(z)$ derivative with respect to $z$ numerically or derived from the reconstructed $f\sigma_8(z)$ under the relation of Eq. \eqref{eq:fprime}. Two approaches give the same results, it is confirmed by checking.  

\rv{Here it would be useful to present the optimized hyperparameters during Gaussian process. The optimized hyperparameters $\sigma_g$ and $\ell$ for reconstructing $D_C(z)$, $D'_C(z)$ and $D''_C(z)$ are $\sigma_{D_C}=2641.43, \ell_{D_C}=1.75$, 
$\sigma_{D'_C}=2642.58, \ell_{D'_C}=1.75$ and $\sigma_{D''_C}=2643.72, \ell_{D''_C}=1.75$ respectively. Similarly, the optimized hyperparameters $\sigma_g$ and $\ell$ for reconstructing $f\sigma_8$ and  $(f\sigma_8(z))'$ are $\sigma_{f\sigma_8}=0.30, \ell_{f\sigma_8}=2.12 $ and $\sigma_{(f\sigma_8)'}=0.30, \ell_{(f\sigma_8)'}=2.12$ respectively.}

At the last step, the dimensionless matter energy density defined as    
 \be
 \Omega_m(z)\equiv\frac{8\pi G\rho_m}{3H^2}=\frac{\Omega_{m0}(1+z)^3}{E^2(z)},
 \ee
 can be expressed in terms the reconstructed function $H(z)=c/D'(z)$, where $E\equiv H/H_0$ is the dimensionless Hubble parameter and cosmological model free.  
 
Finally, substituting the above corresponding quantities into Eq. \eqref{eq:mu} and deriving the uncertainty by error propagation equation, one obtains the reconstructed modified factor $\mu$ with $1-3\sigma$ regions in Figure \ref{fig:fmu}. It can be seen that the modified factor $\mu$ evolves with respect to the redshift $z$, and almost deviates from the standard GR case in $3\sigma$ regions at low redshifts, where the GR $\mu\equiv 1$ case is plotted as red dashed horizontal line for benchmark. In the Figure \ref{fig:fmu} we also plot the $\mu$ factor for the DGP braneworld model as green thin line, and the parameterized form \cite{ref:NesserisGeff2017}  
\be
\mu_P(z)=1+g_a(\frac{z}{1+z})^n-g_a(\frac{z}{1+z})^{2n},
\ee   
with the best fitted values of $n=1$, $g_a=-0.944$ (and $n=2$, $g_a=-1.156$) case denoted as black dash-dot line (and blue dotted line). Here the $f(R)$ gravity case is not shown for the lack of a scale $k$ dependent modified factor $\mu(z,k)$.

\rv{At the last of this section, we want to warn the reader that the functions of $H(z)$, $H'(z)$, $f\sigma_8(z)$ and $[f\sigma_8(z)]'$ are reconstructed from cosmic observational data points without any assumption of cosmological models and any specific values of $\sigma_{8,0}$, but in the last step to obtain $\mu(z)$ via Eq. (\ref{eq:mu}) the values of $\Omega_{m0}$ and $\sigma_{8,0}$ and their uncertainties have to be adopted. Therefore $\Omega_{m0}=0.334\pm0.018$ from SH0ES and $\sigma_{8,0}=0.8111\pm 0.0060$ from {\it Planck} 2018 are adopted, where the uncertainties are included in the final function $\mu$ by error propagation. In this sense, it would be not completely model-independent. One may concern the impacts to the function $\mu$ by taking different values of  $\Omega_{m0}$ and $\sigma_{8,0}$. Actually, from Eq. (\ref{eq:mu}), one can easily see that the value of $\Omega_{m0}$ rescales the reconstructed function $\mu$ as $1/\Omega_{m0}$. Thus the impact to the reconstructed function $\mu$ due to the change of $\Omega_{m0}$ values is simple, i.e. the smaller values of $\Omega_{m0}$ will increase the $\mu$ values. To study the effects to $\mu$ by taking different values of $\sigma_{8,0}$, one can substitute Eq. (\ref{eq3}) and Eq. (\ref{eq:deltaodelta0}) into Eq. (\ref{eq:fdelta}), and then has $f(z)$ and $f'(z)$ as
\ba
f(z)&=&\frac{f\sigma_8(z)}{\sigma_{8,0}-\int_0^z
 \frac{f\sigma_8(\tilde{z})}{1+\tilde{z}}\,d\tilde{z}},\\
 f'(z)&=&\frac{[f\sigma_8(z)]'}{\sigma_{8,0}-\int_0^z
 \frac{f\sigma_8(\tilde{z})}{1+\tilde{z}}\,d\tilde{z}}+\frac{[f\sigma_8(z)]^2}{(1+z)\left[\sigma_{8,0}-\int_0^z
 \frac{f\sigma_8(\tilde{z})}{1+\tilde{z}}\,d\tilde{z}\right]^2}.
\ea
Thus from the above two equations of $f(z)$ and $f'(z)$, one can see that the larger values of $\sigma_{8,0}$ will suppress final function $\mu$ while the cosmic observations reconstructed terms $f\sigma_8(z)$ and $H'/H$ are kept untouched. One can check this effect to $\mu$ when different values of $\sigma_{8,0}$ are adopted in Figure \ref{fig:fmu}. In any situation, the simple $\Lambda$CDM in GR, DGP and parameterized form $\mu_P$ can not match the reconstructed $\mu$ well even in $3\sigma$ regions. It implies that more complicated modified gravity theory is required.  
\begin{figure*}[tbp]  
\includegraphics[scale=0.31]{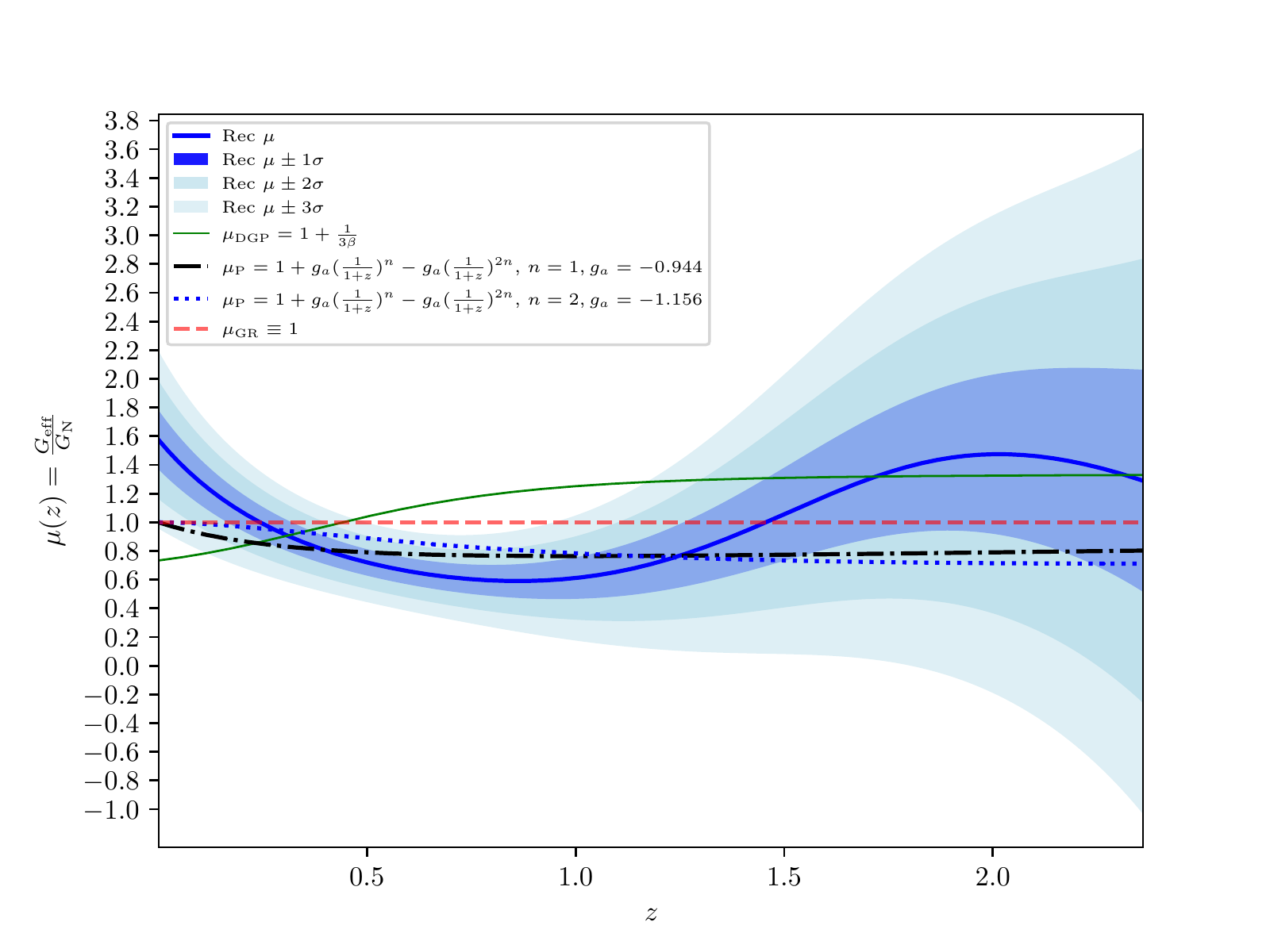}
\includegraphics[scale=0.31]{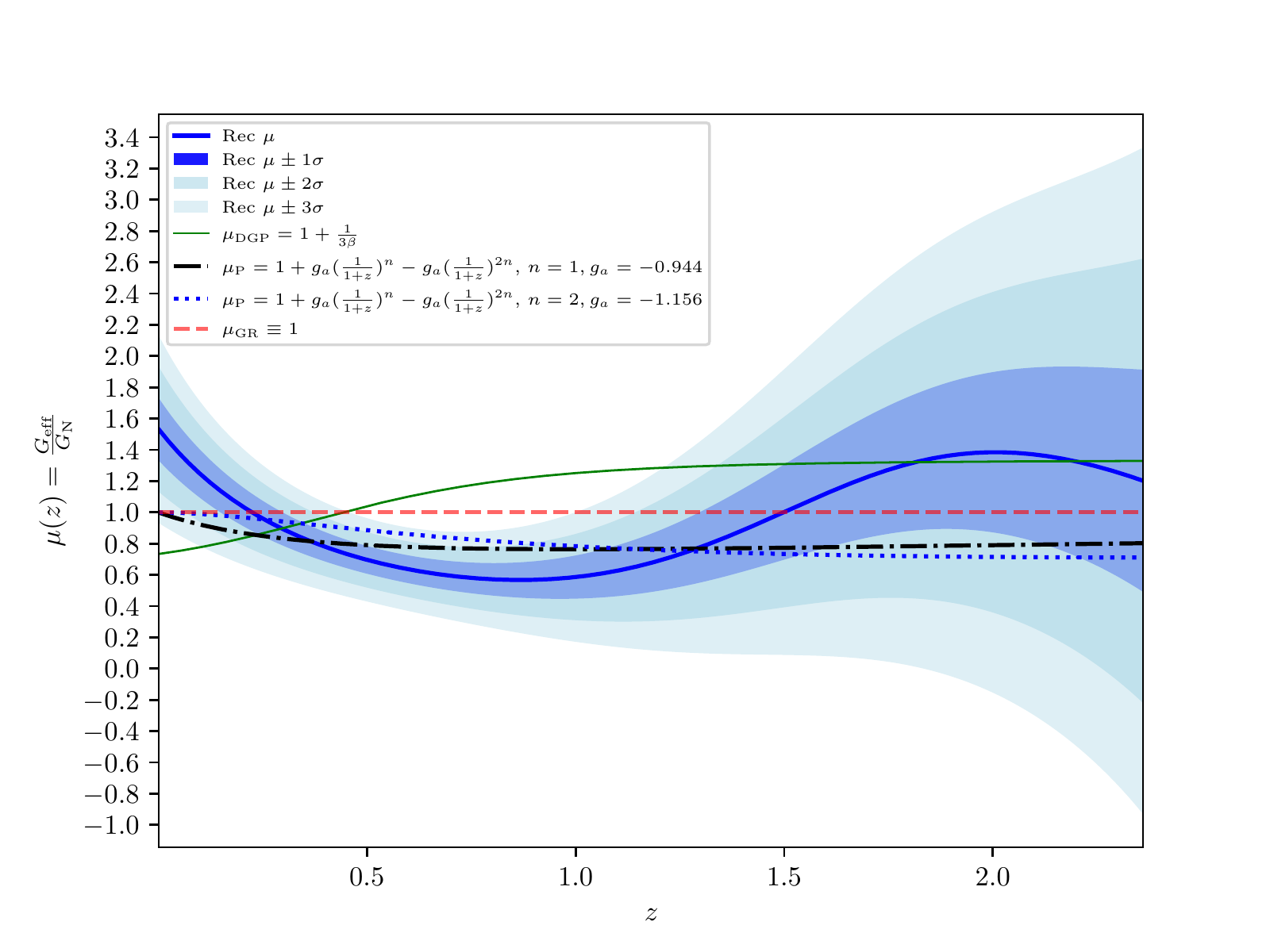}
\includegraphics[scale=0.31]{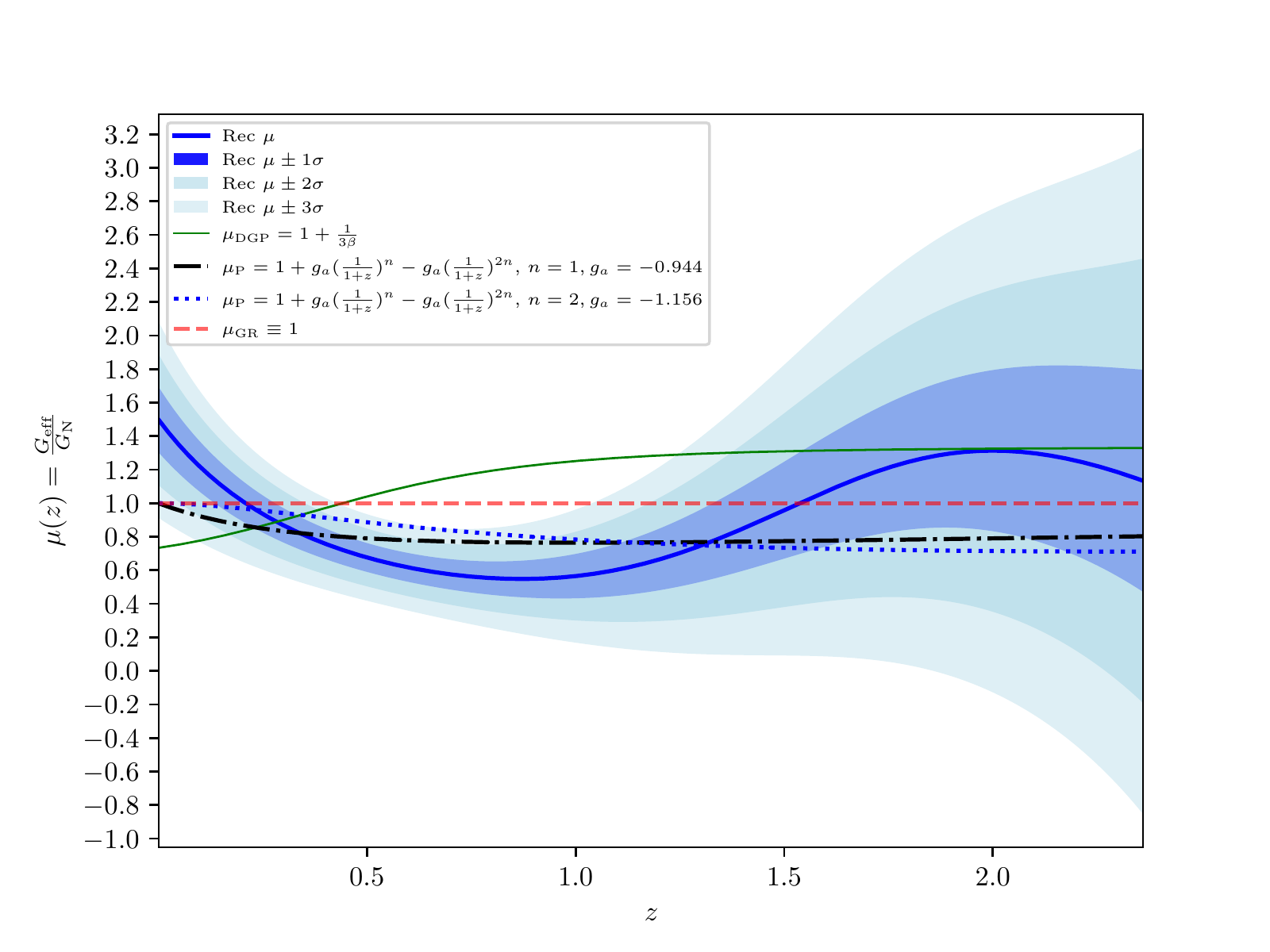}
\caption{The reconstructed modified factor $\mu(z)$ with $1-3\sigma$ shadow regions by taking different values of $\sigma_{8,0}=0.79,0.8111,0.83$ from the left panel to the right panel respectively while $\Omega_{m0}=0.334\pm0.018$ is adopted, where the GR $\mu\equiv 1$ is plotted as red dashed horizontal line for benchmark. The DGP braneworld model $\mu_{\rm DGP}$ is denoted as green thin line, and the parameterized $\mu_P(z)=1+g_a(\frac{z}{1+z})^n-g_a(\frac{z}{1+z})^{2n}$ with best fitted values of $n=1$, $g_a=-0.944$ (and $n=2$, $g_a=-1.156$) is also plotted as black dash-dot line (and blue dotted line).} \label{fig:fmu}
\end{figure*} 
}

\section{Conclusion and Discussion} \label{sec:con}

In this paper we reconstruct a modified factor $\mu$ in a data-driven and almost model-independent way via Gaussian process by the currently available cosmic observations, which consist the Pantheon+ SNe Ia samples, the observed Hubble parameter $H(z)$ and the observed $f\sigma_{8,\rm obs}$ data points. The main result is shown in Figure \ref{fig:fmu}, where a time or redshift varying $\mu$ with $1-3\sigma$ shadow regions is presented. Meanwhile, the evolutions of $\mu$ for the $\Lambda$CDM in GR, DGP and parameterized form $\mu_P(z)=1+g_a(\frac{z}{1+z})^n-g_a(\frac{z}{1+z})^{2n}$ with the best fitted model parameters are also plotted in the Figure \ref{fig:fmu} for comparisons. It shows that none of them can match the reconstructed $\mu$ well whthin $3\sigma$ regions. It implies more complicated modified gravity theories would be favored by cosmic observations.          

In the literatures, another possible parameterized modification to GR, i.e the so-called the 'slip' of gravity potentials is also considered \cite{ref:Amendolajcap2008,ref:BZ2008,ref:Pogosianprd2010,ref:MGCAMB,ref:XuMGGW2015},
\ba
k^2\Psi &=& -4\pi G_{\rm N}\mu(a,k)a^2\rho\Delta,\\
k^2(\Phi+\Psi)&=&-8\pi G_{\rm N}\Sigma(a,k)a^2\rho\Delta,
\ea
where $\rho$ is the background matter density and $\Delta=\delta+3aHv/k$ is the comoving density contrast. $\Sigma$ is equal to one in $\Lambda$CDM. In Refs. \cite{ref:ZhaoPCAprl2009,ref:Hojjatiprd2012} by the principal component analysis (PCA), the eigenmodes and eigenvalues of these functions are found. Based on Horndeski gravity \cite{ref:Horndeski1974,ref:DGSZ2011,ref:Kobayashi2011} in the effective theory (EFT) framework, the $\mu(a,k)$ and $\Sigma(a,k)$ are parameterized as \cite{ref:Silvestri2013,ref:Pogosian2016,ref:EspejoHorndeskiprd2019} 
\ba
\mu(a,k)&=&\frac{m_{0}^{2}}{M_{*}^{2}}\frac{1+M^{2}a^{2}/k^{2}}{f_{3}/2f_{1}M_{*}^{2}+M^{2}(1+\alpha_{T})^{-1}a^{2}/k^{2}},\label{eq:mu_qsa}\\
\Sigma(a,k)&=&\frac{m_{0}^{2}}{2M_{*}^{2}}\frac{1+f_{5}/f_{1}+M^{2}\left[1+(1+\alpha_{T})^{-1}\right]a^{2}/k^{2}}{f_{3}/2f_{1}M_{*}^{2}+M^{2}(1+\alpha_{T})^{-1}a^{2}/k^{2}},
\label{eq:sigma_qsa}
\ea
where $m^{-2}_0=8\pi G_{\rm N}$, $\alpha_T=c^2_T-1$, and $M$, $M_*$, $f_1$, $f_3$ and $f_5$ are dimensionful EFT functions. Here $c_T$ is the speed of gravity. \rv{Based on these parameterized form in Ref. \cite{ref:EspejoHorndeskiprd2019}, $\mu$ and $\Sigma$ are computed at fixed scales $k\in\{0.01,0.085,0.15\} h \rm{Mpc}^{-1}$, where any significant dependence on $k$ was not found. Thus, the functions $\mu(a,k)$ and $\Sigma(a,k)$ are reconstructed at the fixed $k=0.01h\rm{Mpc}^{-1}$ scale in the quasi-static approximation. In Ref. \cite{ref:PogosianNA2022} instead of giving any concrete parameterized forms of $\mu$ and $\Sigma$ such as Eq. (\ref{eq:mu_qsa}) and Eq. (\ref{eq:sigma_qsa}), $\mu$ and $\Sigma$ are parameterized in terms of their values at $11$ discrete values (nodes) of the scale factor $a$ ranging in $[1,0.25]$ with a cubic spline connecting the nodes. The correlation introduced by these $11$ scale factor nodes is eliminated by supplementing a smoothness scale imposed by a theoretical correlation prior. It is quite similar to the kernel function in Gaussian process. The reader should also notice that since we are using the cosmic observations $f\sigma_8(z)$ from RSD, the potential $\Psi$ felt by non-relativistic matter determining the peculiar velocity of galaxies can be constrained. So in this work, we only consider the $k$ dependence free modified factor $\mu$ in stead of the 'slip' factor $\Sigma$. It would be quite safe.}

If the modified factor $\mu$ really describes the ratio of $G_{\rm eff}/G_{\rm N}$, from the Figure \ref{fig:fmu}, one can see a deviation from $G_{\rm N}$ nearly beyond $3\sigma$ at present. Naively, If one moves the whole $\mu$ evolution curves to the position having $G_{\rm eff}/G_{\rm N}=1$, one will still find $\mu_{\rm GR}$ will be above the reconstructed $\mu$ beyond $3\sigma$. In any cases, the deviation from GR cannot be eliminated in $3\sigma$ regions.       

Although in this work, we did not provide a concrete viable action form of modified gravity, along the line of this work, one may reconstruct a concrete form of modified gravity from the reconstructed form of $\mu(z)$ based on Eq. \eqref{eq:mufT}, for example the $f(T)$ form and so on easily, but may not for $f(R)$ form for all $k$ scales due to the scale-dependence is outside the range probed by large scale structure surveys within the linear perturbation theory, and the lack of viable cosmic observations at different scales currently. Of course, one can treat the growth ration evolution function is respected at a fixed scale $k$, say at the scale $k=0.01h\rm{Mpc}^{-1}$ in the quasistatic approximation where the linear perturbation theory holds well \cite{ref:EspejoHorndeskiprd2019}, then the form of $f(R)$ can be also reconstructed at some peculiar scales $k$ by repeating our process as required. And one can also investigate the growth index via the definition $f(z)=\Omega^\gamma_{m}(z)$ easily, since we already have the reconstructed growth function $f(z)$ and $\Omega_{m}(z)$ in hand. Actually, one can also reconstruct the gravitational potentials 'slip' based on the methodology proposed in this work. We will leave them in the future work. 

\begin{acknowledgments}
L. Xu's work is supported in part by National Natural Science Foundation of China under Grant No. 12075042 and No. 11675032. And E.-K. Li's work supported in part by Natural Science Foundation of Guangdong Province of China under Grant No. 2022A1515011862. 
\end{acknowledgments}


\begin{thebibliography}{99}

\bibitem{ref:Riess98} A. G. Riess, \etal, Astron. J. 116, 1009 (1998) [astro-ph/9805201].

\bibitem{ref:Perlmuter99} S. Perlmutter, \etal, Astrophys. J. 517, 565(1999) [astro-ph/9812133].

\bibitem{ref:DEReview1} S. Weinberg, Rev. Mod. Phys. 61 1(1989).

\bibitem{ref:DEReview2} V. Sahni and A. A. Starobinsky, Int. J. Mod. Phys. D 9 373(2000) [arXiv:astro-ph/9904398].

\bibitem{ref:DEReview3} S. M. Carroll, Living Rev. Rel. 4 1(2001) [arXiv:astro-ph/0004075].

\bibitem{ref:DEReview4} P. J. E. Peebles, B. Ratra, Rev. Mod. Phys. 75 559(2003) [arXiv:astro-ph/0207347].

\bibitem{ref:DEReview5} T. Padmanabhan, Phys. Rept. 380 235(2003) [arXiv:hep-th/0212290].

\bibitem{ref:DEReview6} E. J. Copeland, M. Sami and S. Tsujikawa, Int. J. Mod. Phys.
D 15 1753(2006) [arXiv:hep-th/0603057].

\bibitem{ref:DEReview7}  M. Li, X. D. Li, S. Wang, Y. Wang, Commun. Theor. Phys. 56, 525(2011), arXiv:1103.5870 [astro-ph.CO].

\bibitem{ref:MG1} S. Tsujikawa, Lect. Notes Phys. 800, 99 (2010), arXiv:1101.0191 [gr-qc].

\bibitem{ref:MG2} T. Clifton, P. G. Ferreira, A. Padilla, C. Skordis, Phys. Rep. 513, 1 (2012), arXiv:1106.2476 [astro-ph.CO].    

\bibitem{ref:MG3}  A.De Felice, S. Tsujikawa, Living Rev. Relativity, 13, 3(2010).

\bibitem{Ishak:2018his} M.~Ishak, Living Rev.\ Rel.\  22, 1 (2019) [arXiv:1806.10122].

\bibitem{Amendola:2016saw} L.~Amendola \etal, Living Rev.\ Rel.\  21, 1(2018) [arXiv:1606.00180].

\bibitem{Joyce:2014kja} A.~Joyce, B.~Jain, J.~Khoury and M.~Trodden, Phys.\ Rept.\  568, 1 (2015) [arXiv:1407.0059].

\bibitem{Nojiri:2010wj} S.~Nojiri and S.~D.~Odintsov, Phys. Rept. 505, 59-144(2011), [arXiv:1011.0544 [gr-qc]].

\bibitem{Nojiri:2017ncd} S.~Nojiri, S.~D.~Odintsov and V.~K.~Oikonomou, Phys. Rept. 692, 1-104(2017), [arXiv:1705.11098 [gr-qc]].
  
 \bibitem{ref:background} S. Capozziello, V. F. Cardone, and A. Troisi, Phys. Rev. D 71, 043503 (2005); S. Nojiri and S.D. Odintsov, Phys. Rev. D 74, 086005 (2006); Y.S. Song, W. Hu, and I. Sawicki, Phys. Rev. D 75, 044004 (2007).  

\bibitem{Peebles} P. J. E. Peebles, {\it The Large-Scale Structure of the
Universe} (Princeton University Press, Princeton, New Jersey 1980).

\bibitem{Fry} J. N. Fry, Phys. Lett. B 158, 211 (1985).

\bibitem{Lightman} A. P. Lightman, P. L. Schechter, Astrophys. J.
{\bf 74}, 831 (1990).

\bibitem{wang} L. Wang, P. J, Steinhardt, Astro. Phys. J. 508, 483(1998).

\bibitem{lue}  A. Lue, R. Scoccimarro, G. Starkman, Phys. Rev. D 69, 124015 (2004)

\bibitem{Aquaviva}  V. Acquaviva, A. Hajian, D.N. Spergel and S. Das, Phys. Rev. D 78, 043514 (2008).

\bibitem{gong08b} Y. G. Gong, Phys. Rev. D 78, 123010 (2008).

\bibitem{polarski} D. Polarski, R. Gannouji, Phys. Lett. B 660, 439 (2008).

\bibitem{Koyama} K. Koyama, R. Maartens, J. Cosmol. Astropart. Phys. 01 (2006) 016.

\bibitem{Koivisto} T. Koivisto, D.F. Mota, Phys. Rev. D 73, 083502 (2006).

\bibitem{Daniel} S. Daniel, R. Caldwell, A. Cooray and A. Melchiorri, Phys. Rev. D 77, 103513 (2008).

\bibitem{knox} L. Knox, Y.-S. Song, J. A. Tyson, Phys. Rev. D 74, 023512 (2006).

\bibitem{ishak2006} M. Ishak, A. Upadhye and D. N. Spergel, Phys. Rev. D 74, 043513 (2006).

\bibitem{linder} E. V. Linder, Phys. Rev. D 72, 043529 (2005).

\bibitem{laszlo} I. Laszlo, R. Bean, Phys. Rev. D 77, 024048 (2008).

\bibitem{Zhang} B. Jain, P. Zhang, Phys. Rev. D78, 063503, (2008).

\bibitem{Hu} W. Hu, I. Sawicki, Phys. Rev. D76, 104043, (2007).

\bibitem{ref:Ishak} M. Ishak, J. Dosset, Phys. Rev. D 80, 043004(2009).

\bibitem{ref:Xu2013} L. Xu, Phys. Rev. D88, 084032, (2013). 

\bibitem{ref:Xuprd2015} L. Xu, Phys. Rev. D91, 063008, (2015). 

\bibitem{ref:DGP} G. R. Dvali, G. Gabadadze, M. Porrati, Phys. Lett. B, 485, 208–214 (2000) [hep-th/0005016].

\bibitem{ref:ZhengHuangfT2011} R. Zheng, Q.-G. Huang, JCAP03(2011)002.

\bibitem{ref:Amendolajcap2008} L. Amendola, M. Kunz, and D. Sapone, JCAP 04,013(2008) .

\bibitem{ref:BZ2008} E. Bertschinger and P. Zukin, Phys. Rev. D 78, 024015 (2008).

\bibitem{ref:Pogosianprd2010} L. Pogosian, A. Silvestri, K. Koyama, and G.-B. Zhao, Phys. Rev. D 81, 104023 (2010).

\bibitem{ref:MGCAMB} A. Hojjati, L. Pogosian, G.-B. Zhao, JCAP 1108,005 (2011). 

\bibitem{ref:XuMGGW2015} L. Xu, Phys. Rev. D91, 103520, (2015). 

\bibitem{ref:FeliceKobayashiTsujikawa2011} A. de Felice, T. Kobayashi, and S. Tsujikawa, Phys. Lett. B 706, 123 (2011), arXiv:1108.4242 [gr-qc].

\bibitem{ref:Solomon2014} A. R. Solomon, Y. Akrami, and T. S. Koivisto, arXiv:1404.4061.

\bibitem{ref:Konnig2014} F. Konnig, Y. Akrami, L. Amendola, M. Motta, and A. R. Solomon,arXiv:1407.4331.

\bibitem{ref:weakness} S. Nesseris, L. Perivolaropoulos, Phys. Rev. D 77, 023504 (2008); S. Basilakos, arXiv:1202.1637 [astr-ph.CO].

\bibitem{ref:Song} Y.-S. Song, W. J. Percival, JCAP, 10, 4(2009).

\bibitem{ref:Li2021} E.-K. Li, M. Du, Z.-H. Zhou, H. Zhang and L. Xu, Mont. Not. Roy. Astro. Soc. 501,4452–4463(2021).

 \bibitem{Alcock:1979mp} C. Alcock, B. Paczynski, Nature, 281, 358(1979).

\bibitem{Scolnic:2021} D. M. Scolnic \etal,  [arXiv:2112.03863 [astro-ph.CO]].
  
\bibitem{Brout:2022} D. Brout \etal, [arXiv:2202.04077 [astro-ph.CO]].
  
\bibitem{Rasmussen:2006} C.~E.~Rasmussen and C.~K.~I.~Williams, {\it Gaussian Processes for Machine Learning}, MIT Press (2006).
  
\bibitem{GaPP:2012} M. Seikel, C. Clarkson, and M. Smith, JCAP06, 036(2012).  
  
\bibitem{ref:Wei2019} Z.-Y. Yin, H. Wei, Sci. China-Phys. Mech. Astron. 62, 999811 (2019).  

\bibitem{Holsclaw:2010prl} T. Holsclaw \etal, Phys. Rev. Lett. 105, 241302 (2010).

\bibitem{Holsclaw:2010prd} T. Holsclaw \etal, Phys. Rev. D 82, 103502 (2010).

\bibitem{Santos-da-Costa:2015} S. Santos-da-Costa \etal, JCAP 10, 061 (2015).

\bibitem{Shafieloo:2012} A. Shafieloo \etal, Phys. Rev. D 85, 123530(2012).

\bibitem{Yahya:2014} S. Yahya \etal, Phys. Rev. D 89, 023503(2014).

\bibitem{Yang:2015} T. Yang \etal, Phys. Rev. D 91, 123533 (2015).

\bibitem{Cai:2016} R. G. Cai \etal, Phys. Rev. D 93, 043517 (2016).

\bibitem{Zhang:2016} M. J. Zhang, J. Q. Xia, JCAP 12, 005(2016), [arXiv: 1606.04398].

\bibitem{Cai:2015} R. G. Cai \etal, JCAP 08, 016 (2016).

\bibitem{Wang:2017} D. Wang, X.-H. Meng, Phy. Rev. D, 023508 (2017).

\bibitem{ref:Liang2022} N. Liang, Z. Li, X. Xie, P. Wu, Astrophys. J.  941,84(2022). 

\bibitem{ref:Mu2023} Y. Mu, B. Chang, L. Xu, arXiv:2302.02559 [astro-ph.CO]. 

\bibitem{ref:gapptelep2021} J. L. Said, J. Mifsud, J. Sultana, K. Z. Adami, JCAP06(2021)015, arXiv:2103.05021 [astro-ph.CO]. 

\bibitem{ref:gappH0Colgain2021} E. O. Colgain, M. M. Sheikh-Jabbari, arXiv:2101.08565 [astro-ph.CO]. 

\bibitem{ref:gappCalderon2022} R. Calderon, B. L'Huillier, D. Polarski, A. Shafieloo, A. A. Starobinsky, arXiv:2206.13820 [astro-ph.CO].

\bibitem{ref:gappLi2022} Z. Li, B. Zhang, N. Liang, arXiv:2212.14291 [astro-ph.CO].

\bibitem{ref:gappCalderon2023} R. Calderon, B. L'Huillier, D. Polarski, A. Shafieloo, A. A. Starobinsky, arXiv:2301.00640 [astro-ph.CO].

 \bibitem{ref:cosmography2011} L. Xu, Y. Wang, Phys. Lett. B, 702, 114(2011).

\bibitem{Aghanim:2018eyx} N.~Aghanim {\it et al.} [Planck Collaboration], arXiv:1807.06209 [astro-ph.CO].

\bibitem{Akrami:2018vks} Y.~Akrami {\it et al.} [Planck Collaboration], arXiv:1807.06205 [astro-ph.CO].

\bibitem{Macaulay:2013swa} E. Macaulay, I.~K. Wehus, H.~K. Eriksen, Phys. Rev. Lett. 111, 161301(2013).
 
\bibitem{Kazantzidis:2018rnb} L. Kazantzidis, L. Perivolaropoulos, Phys. Rev. D 97, 103503(2018).
 
\bibitem{Riess:2022} A. G. Riess \etal, Astrophys. J. Lett. 934, L7(2022).
 
\bibitem{ref:NesserisGeff2017} S. Nesseris, G, Pantazis, L. Perivolaropoulos, Phy. Rev. D 96, 023542 (2017). 
 
\bibitem{ref:ZhaoPCAprl2009} G.-B. Zhao, L. Pogosian, A. Silvestri, J. Zylberberg, Phy. Rev. Lett. 103, 241301 (2009).

\bibitem{ref:Hojjatiprd2012} A. Hojjati, G.-B. Zhao, L. Pogosian, A. Silvestri, R. Crittenden, K. Koyama, Phys. Rev. D 85, 043508 (2012).
 
\bibitem{ref:Horndeski1974} G. W. Horndeski, Int. J. Theor. Phys. 10, 363 (1974).
 
\bibitem{ref:DGSZ2011} C. Deffayet, X. Gao, D. A. Steer, and G. Zahariade, Phys. Rev. D 84, 064039 (2011).
 
\bibitem{ref:Kobayashi2011} T. Kobayashi, M. Yamaguchi, and J. Yokoyama, Prog. Theor. Phys. 126, 511 (2011).
 
\bibitem{ref:Silvestri2013} A. Silvestri, L. Pogosian, and R. V. Buniy, Phys. Rev. D 87, 104015 (2013).
 
\bibitem{ref:Pogosian2016}  L. Pogosian and A. Silvestri, Phys. Rev. D 94, 104014 (2016).
 
\bibitem{ref:EspejoHorndeskiprd2019} J. Espejo, S. Peirone, M. Raveri, K. Koyama, L. Pogosian, A. Silvestri, Phy. Rev. D 99, 023512 (2019).
  
\bibitem{ref:PogosianNA2022} L. Pogosian, M. Raveri, K. Koyama, M. Martinelli, A. Silvestri, G.-B. Zhao, J. Li, S. Peirone, A. Zucca, Nat. Astro. 6, 1484(2022).

\end{thebibliography}
\end{document}